\documentclass[10pt,aps,preprint,pra,twocolumn]{revtex4}
\usepackage{amsmath}
\usepackage{graphicx}
\usepackage{amssymb}
\usepackage{multirow}
\usepackage{color}
\newcommand{\upperRomannumeral}[1]{\uppercase\expandafter{\romannumeral#1}}

\begin{document}

\title{First principle and tight-binding study of strained SnC}
\author{Y. Mogulkoc}
\affiliation{Department of Engineering Physics, Faculty of Engineering, Ankara University, 06100, Tandogan, Ankara, Turkey}
    
\author{M. Modarresi}
\affiliation{Department of Physics, Ferdowsi University of Mashhad, Mashhad, Iran.}

\affiliation{Laboratory of Organic Electronics, Department of Science and Technology, Campus Norrk\"oping, Link\"oping University, SE-60174 Norrk\"oping, Sweden}

\author{A. Mogulkoc}
\email{mogulkoc@science.ankara.edu.tr}
\affiliation{Department of Physics, Faculty of Sciences, Ankara University, 06100, Tandogan, Ankara, Turkey}
\author{Y. O. Ciftci}
\affiliation{Department of Physics, Faculty of Sciences, Gazi University, 06500, Teknikokullar, Ankara, Turkey}
\author{B. Alkan}
\affiliation{Department of Engineering Physics, Faculty of Engineering, Ankara University, 06100, Tandogan, Ankara, Turkey }
\date{\today}

\begin{abstract}
We study the electronic and optical properties of strained single-layer SnC in the density functional theory (DFT) and tight-binding models. We extract the hopping parameters tight-binding Hamiltonian for monolayer SnC by considering the DFT results as a reference point. We also examine the phonon spectra in the scheme of DFT, and analyze the bonding character by using Mulliken bond population. Moreover, we show that the band gap modulation and transition from indirect to direct band gap in the compressive strained SnC. The applied tensile strain reduces the band gap and eventually the semiconductor to semimetal transition occurs for 7.5\% of tensile strain. In the framework of tight-binding model, the effect of spin-orbit coupling on energy spectrum are also discussed. We indicate that while tensile strain closes the band gap, spin-orbit gap is still present which is order of $\sim 40$ meV at the $\Gamma$ point. The substrate effect is modeled through a staggered sub-lattice potential in the tight-binding approximation. The optical properties of pristine and strained SnC are also examined in the DFT scheme. We present the modulation of real and imaginary parts of dielectric function under applied strain.  

\end{abstract}

\maketitle
\section{Introduction}
2D materials are promising candidates for materials science and technology. Among these, graphene shows interesting electronic and transport properties but its technological applications are limited due to the negligible electronic band gap at the $\mathrm{K}$ point. Since the band gap at the Fermi level is essential for controlling the conductivity in electronic devices, many studies have been made in order to open up the energy band gap of graphene \cite{GrGAP1,GrGAP2,GrGAP3,GrGAP4}. Contrary to the graphene, some group \upperRomannumeral{4} graphene-like 2D structures, i.e., silicene \cite{Si1,PSSR:PSSR201510338}, germanene \cite{Ge1,GeSn,PSSR:PSSR201510338}, stanene \cite{Sn1,Sn2,GeSn} and
their binary compounds \cite{SnC1,Binary1,C2JM30915G,Binary2,Binary3,Binary4} have energy band gap which enable the possible technological applications. Even graphene has gapless band structure, stanene has band gap, $E_{g}$, around $\sim 75$ meV \cite{GeSn} due to the its relatively higher spin-orbit coupling (SOC). On the other hand, previous theoretical studies confirm that the applied strain modifies the electronic properties of group \upperRomannumeral{4} binary compounds \cite{C2JM30915G,Binary4}. Among these, monolayer honeycomb lattice of SnC (two dimensional binary compound of carbon and tin atom) shows planar structure where atoms are located on the same plane with indirect energy band gap between $\mathrm{\Gamma}$ and $\mathrm{K}$ high symmetry points with lower SOC according to stanene. Even there is no experimental work related this material, some theoretical calculations were devoted to electronic, elastic and optical properties of bulk \cite{:/content/aip/journal/jap/88/11/10.1063/1.1287225,:/content/aip/journal/jap/108/6/10.1063/1.3478717} and monolayer \cite{C2JM30915G,PhysRevB.80.155453,SnC1} SnC. An indirect-direct band gap transition of SnC was predicted through adatom decoration \cite{SnC1} and applied strain \cite{C2JM30915G} in the scheme of DFT.

From an experimental point of view,  quite recently a novel technique for synthesis of 2D group-IV binary compounds is presented \cite{1347-4065-56-5S1-05DA06}. Recently, there are also some reports on the synthesis of GeSn \cite{1468-6996-16-4-043502,Taoka201748,1347-4065-56-1S-01AB05} and SiGe \cite{1347-4065-55-3S1-03CB01,1347-4065-42-4S-1933,1347-4065-56-5S1-05DA06}. However, there is no experimental report on the synthesis of SnC.

In this paper, we have reported that the strain modulation of electronic structure in the framework of density functional theory and tight-binding model. Even the evolution of band structure with applied strain was examined in Ref. \onlinecite{C2JM30915G}, we have constructed a tight-binding model together with spin-orbit coupling. For the hexagonal crystal structure of SnC, we have investigated not only electronic structure but also vibrational properties to complete the whole study for stability condition. Moreover, the optical properties can provide detailed information about the electronic structure of the materials. The optical properties of solids are a major topic, both in basic research and industrial applications. While for the former the origin and nature of different excitation processes is of fundamental interest, the latter can make use of them in many optoelectronic devices. Through this motivation, we have also studied the optical properties of pristine and strained SnC in the framework of DFT. The technological interest in SnC is driven by its elasticity, electronic and optical properties that makes it appropriate for optoelectronic devices.  

\section{Model and Method}

\subsection{Density Functional Theory}
We examine the electronic properties and phonon spectrum of monolayer SnC in the scheme of DFT. All of the first-principle calculations are performed by using the VASP package \cite{PhysRevB.54.11169} except the charge density and atomic population analysis are achieved by CASTEP package \cite{clark2005first}. The exchange-correlation potential is approximated by generalized gradient approximation (GGA) with the Perdew-Burke-Ernzerhof (PBE) functional \cite{PhysRevLett.77.3865,PhysRevLett.78.1396}. A plane-wave basis set with kinetic energy cut-off of 500 eV is considered while performing all first-principle calculations. Atomic positions and lattice constant are optimized by conjugate gradient method. Furthermore, Brillouin zone integration is performed by using the Monkhorst-Pack method \cite{PhysRevB.13.5188} of $24\times24\times1$ $\mathrm{k}$-points with vacuum spacing of $30$ \AA{} set along the perpendicular direction to the monolayer SnC. In the geometrical optimization, the maximum Hellmann-Feynman force acting on each atom is less than $0.001$ eV/\AA{}. In addition to DFT-PBE, hybrid Heyd-Scuseria-Ernzerhof (HSE06) \cite{HSE1,HSE2} is employed as hybrid functional method which considers an exchange-correlation functional consists of Hartree-Fock (HF) exact exchange functional and the PBE functional. Phonon spectrum of SnC is calculated by VASP package with small displacement method using PHONOPY code \cite{phono3py}. A ($4\times4\times1$) supercell consisting of 32 atoms is considered for the phonon calculation to determine the mechanical stability of structure. 

\begin{figure}[!t]
\includegraphics[width=0.7\linewidth]{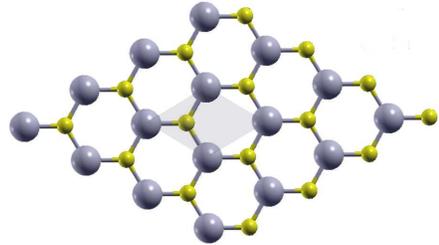}
\caption{The planar honeycomb structure of SnC consists of carbon (yellow) and tin (gray) atoms. The shaded area represents the unit cell of SnC.}\label{FIG1}
\end{figure}

\subsection{Tight-binding calculations}
For our tight-binding calculations we consider the one \textit{s} and three \textit{p} atomic orbitals in carbon and tin atoms. The difference between two atoms in the unit-cell leads to different on-site energy in the tight-binding model. The hopping between atomic orbitals is written in the Slater-Koster model. Eqn.(\ref{1}) represents the total tight-binding Hamiltonian,
\begin{eqnarray}
H=\sum_{i,\alpha} \epsilon_{i,\alpha} c_{i,\alpha}^{\dagger}c_{i,\alpha}+\sum_{<i,j>,\alpha,\beta} t_{i,j,\alpha,\beta}^0 c_{i,\alpha}^{\dagger }c_{j,\beta} \label{1}
\end{eqnarray}

where the $c_{i,\alpha}^{\dagger}$ and $c_{i,\alpha}$ are creation and annihilation operators for electron on the atom $i\in {[C,Sn]}$ and orbital $\alpha\in {[s,p_x,p_y,p_z]}$. We consider four atomic orbital per atom and the total tight-binding Hamiltonian is a $8\times8$ matrix. The required parameters which include the on-site energy $\epsilon_{i,\alpha}$ and hopping between nearest neighbor atoms $<i,j>$ and $\alpha$ and $\beta$ atomic orbitals $t_{i,j,\alpha,\beta}^0$ are obtained by fitting the tight-binding band structure with DFT results. We include the SOC in the tight-binding calculations. The SOC Hamiltonian is proportional to the internal dot product of orbital and spin angular momentums $H^{SOC}=\lambda L.S$, where $\lambda$ is the SOC interaction strength. By using the ladder operators the $H^{SOC}$ is expanded based on the tight-binding basis sets \cite{SOCtheort}. The SOC strength of carbon atoms is negligible and for tin atoms its value is around, $\lambda=0.2 $ eV \cite{GeSn}. In the tight-binding model, the external applied strain modify the hopping integral between carbon and tin atomic orbitals. To plot the band structure, the total Hamiltonian matrix is diagonalized for each $k$ point in a specific direction of first Brillouin zone.

\subsection{Optical properties}
To examine the possible technological application of pristine and strained structure of SnC, we investigated the optical properties by using GGA-PBE functional without local field effect. The linear response of a system due to an external electromagnetic radiation is described by the complex dielectric function $\varepsilon (\omega)$=$\varepsilon_{1} (\omega) +i\varepsilon_{2} (\omega)$ \cite{sun2004optical}. The dispersion of the imaginary part of complex dielectric function $ \varepsilon_{2} (\omega) $ was derived from the momentum matrix elements between the occupied and unoccupied wave functions as follows,

\begin{eqnarray}
\varepsilon_{2}^{(\alpha\beta)}&=&\dfrac{4\pi^{2}e^{2}}{\Omega}\lim_{q\to 0} \dfrac{1}{q^{2}}\sum\limits_{c,v,\boldsymbol{k}}2\omega_{\boldsymbol{k}}\delta\left( \epsilon_{c\boldsymbol{k}}-\epsilon_{v\boldsymbol{k}}-\omega\right)\notag \\
&\times&\left\langle u_{c+\boldsymbol{k}+\boldsymbol{e}_{\alpha q}} \lvert u_{v\boldsymbol{k}} \right\rangle \left\langle u_{c+\boldsymbol{k}+\boldsymbol{e}_{\alpha q}} \lvert u_{v\boldsymbol{k}} \right\rangle^{*} \notag
\end{eqnarray}%,

where the $ c$ and $ v$ correspond to conduction and valence band states respectively, and $ u_{c{\mathbf{k}}}$ is the cell periodic part of the orbitals at the k-point $ \bf k$. The real component of the dielectric function, $\varepsilon_{1} (\omega)$ is calculated via the Kramers$-$Kronig relation \cite{hu2007first}. The dielectric function provides a comprehension of the optical properties as it relates the electronic response of the system to the exposure of electromagnetic radiations. As the hexagonal crystal structure of SnC, they are characterized by two independent tensor components (perpendicular and parallel to $z$ axis) of the dielectric tensor. Here we have investigated the influence of strain on the in-plane component (E $\parallel$ z) of dielectric functions related with strictly 2D structure of SnC.  
  
\section{Results and discussions}

The monolayer SnC consists of two different atoms (carbon and tin) based on planar honeycomb structure as shown in Fig. \ref{FIG1}. The lattice parameter of SnC is calculated as 3.59 \AA{} which is comparable with previous reports \cite{PhysRevB.80.155453,C2JM30915G}. We calculate the electronic band structure of SnC in the PBE, tight-binding and HSE06 models. It is well-known fact that DFT-PBE includes unphysical self-Coulomb repulsion \cite{PhysRevB.23.5048} which yields underestimation of energy band gap \cite{PhysRevLett.51.1884,PhysRevLett.51.1888,PhysRevLett.100.146401}. By inclusion of short-range exact Hartree$-$Fock exchange with HSE06, the Coulomb self-repulsion error can be decreased \cite{PhysRevB.63.144510} that gives rise to better estimation of band gap \cite{:/content/aip/journal/jcp/125/22/10.1063/1.2404663}.

\begin{figure}[!t]
\includegraphics[width=\linewidth]{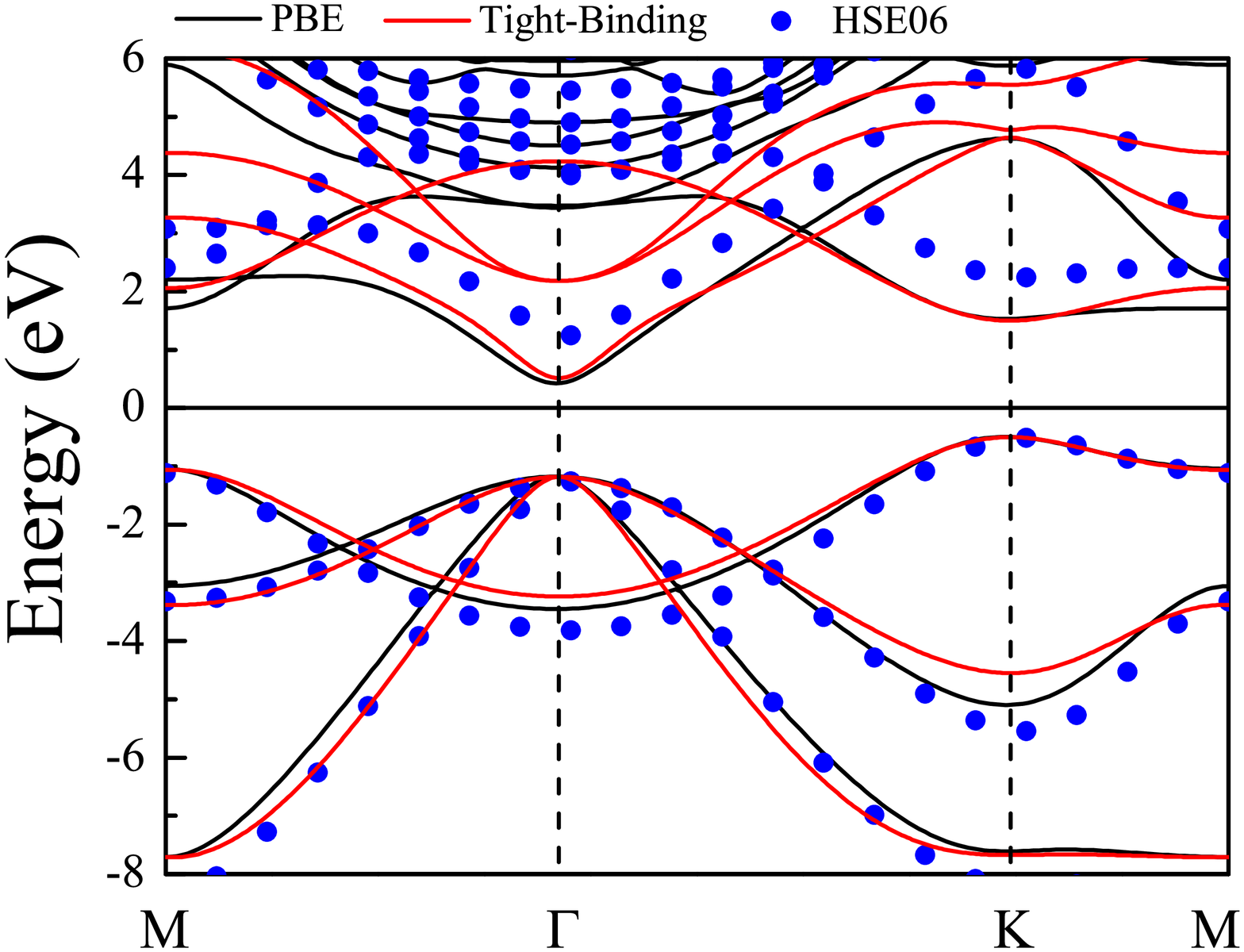}
\caption{The band structure of SnC in the DFT-PBE (black line), tight-binding (red line) and DFT-HSE06 (blue dot).}
\label{BANDS}
\end{figure}

\begin{table}[ht]
\caption {Tight binding parameters for SnC.} \label{tabTB}
\begin{center}
\begin{tabular}{ |c|c|c| } 
 \hline
 Parameter  &   Value ($eV$)\\
 \toprule
 $t_{ss\sigma}$ & -1.6\\ 
 $t_{sp\sigma}$ & 2.9\\ 
 $t_{pp\sigma}$ & 2.1\\ 
 $t_{pp\pi}$ & -1.2\\ 
 $\epsilon_s (C)$ & -6.9\\
 $\epsilon_s (Sn)$ & -1.9\\
 $\epsilon_p (Sn)$ & 2\\
 \hline
\end{tabular}
\end{center}
\end{table}
The on-site energy of \textit{p} atomic orbital for carbon atoms is set to zero as the reference atomic energy. The electronic band structure in three different models is presented in Fig. \ref{BANDS}. 

All three models predict an indirect band gap between the $\Gamma$ and K points in the first Brillouin zone. The Valence Band Maximum (VBM) is at the K point and the Conduction Band Minimum (CBM) is at the $\Gamma$ points of the first Brillouin zone. The size of band gap in DFT is 0.92 eV and 1.75 eV in the HSE06 model. The valence electronic bands are matched in the DFT and HSE06 model. The discrepancy between DFT and HSE06 is related to the conduction bands. To minimized the computational cost of the calculations we use the PBE functional in the following to fit with the tight binding model. The obtained tight-binding parameters are presented in Table \ref{tabTB}. The hopping parameters are much lower than graphene, but hopping signs are in complete agreement with reported parameters of graphene \cite{SOCtheort}. The electronic band gap in the tight binding model is 1.02 eV which is originated from the difference of atomic on-site energy of carbon and tin atoms. The presence of band gap in SnC potentially may solve the zero-gap problem in graphene-based materials.

\begin{figure}[!t]
\includegraphics[width=\linewidth]{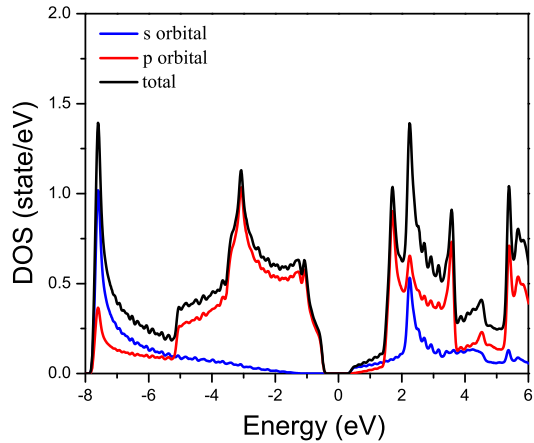}
\caption{The PDOS of SnC in the DFT-PBE .}
\label{PDOS1}
\end{figure}

To determine the contribution of each atomic orbital we calculate the Partial Density Of State (PDOS) within the framework of DFT-PBE. The PDOS in Fig. \ref{PDOS1} reveals that the contribution of \textit{s} and \textit{p} atomic orbitals around the Fermi level. The VBM mostly comes from the atomic \textit{p} orbitals but the CBM in the $\Gamma$ point is related to atomic \textit{s} orbitals. Due to the essential role of atomic \textit{s} orbital, both \textit{s} and \textit{p} orbitals should be taking into account at the tight-binding model. 

\begin{figure}[!t]
\includegraphics[width=0.7\linewidth]{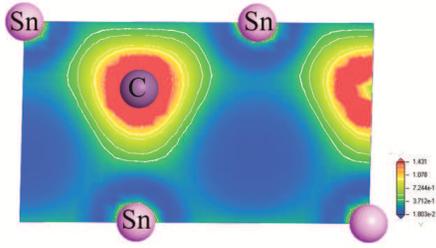}
\caption{Charge density of SnC.}\label{FIG4}
\end{figure}

In order to calculate the charge distribution between the C and Sn atoms, we have performed the Mulliken population analysis scheme. The structure shows an electronic charge state of 1.13 electrons for the Sn atoms, while the C atoms -1.13 (See  Table \ref{tab2}). Calculated bond length is also compatible with those studies which considered local density approximation (LDA) \cite{C2JM30915G} and GGA \cite{PhysRevB.80.155453} exchange-correlation potentials.
\begin{table}[htp]
\caption{Atomic Populations (Mulliken)} \label{tab2}
\begin{tabular}{|c|c|c|c|c|c|}
\hline
Species           & Ion & $s$ & $p$ & Total & Charge (e)\\
\hline
C                 & 1 & 1.52 & 3.60 & 5.13 & -1.13   \\
Sn                & 1 & 1.18 & 1.69 & 2.87 &  1.13   \\
\hline
\end{tabular}
\\

\begin{tabular}{|c|c|c|}
\hline
Bond  \qquad  & Population \qquad  & Length (\AA{}) \\
\hline
C-Sn    & 2.44 \quad&  2.08\\
\hline
\end{tabular}

\end{table}
Additionally, to mention the nature of bonding between the Sn and C atoms we have plotted the orbital and charge density contour plots of the structure of SnC. From the charge density plot of the structure in Fig. \ref{FIG4}, it can be seen that the hybridization occurs between \textit{s} and \textit{p} orbitals of carbon atoms. The Mulliken population analysis can facilitate to assign the electrons in several fractional methods among the various parts of bonds and overlap population has correlations with covalency or ionicity of bonding and bond strength. Population analysis in CASTEP is carried out using a projection of the plane wave states onto a localized basis developed by Ref. \onlinecite{SANCHEZPORTAL1995685}. Population analysis of the resulting projected states is then accomplished by using the Mulliken formalism \cite{:/content/aip/journal/jcp/23/10/10.1063/1.1740588}. The Mulliken charge and overlap population are useful in evaluating the covalent, ionic, or metallic nature of bonds in the system. A high value of the bond population indicates a covalent bond, whilst a low value indicates an ionic nature. If Mulliken population (MP) value is very small (or zero), bond is more ionic (ideal ionic), namely, MP$\rightarrow$0, bond ionicity increases. Also, positive and negative values of the population indicate bonding and anti-bonding states, respectively. Higher positive value of MP indicates a high degree of covalency in the bond. There is one type of bond with positive Mulliken population value in the crystal structure of SnC. Calculated Mulliken bond population is 2.44 which is  lower than graphene (3.05 \cite{john2016theoretical}) and is higher than stanene (1.99 \cite{john2016theoretical}). It can be concluded that SnC is more covalent than stanene. 

\begin{figure}[!t]
\includegraphics[width=0.9\linewidth]{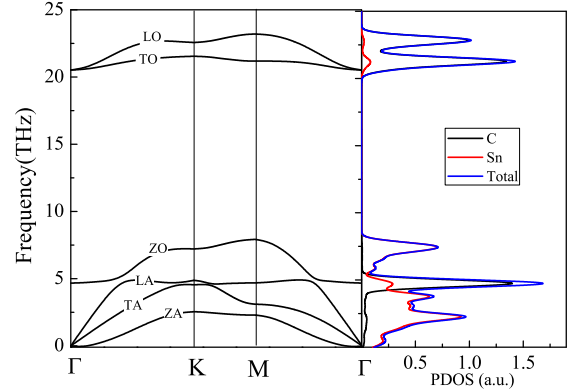}
\caption{Phonon spectrum and partial phonon DOS of SnC.} \label{FIG5}
\end{figure}

The phonon spectrum of SnC is illustrated in Fig. \ref{FIG5}. It is well-known fact that, spectrum consists of six phonon branches due to the $\mathrm{D}_{6h}$ point-group symmetry of the structure. While four of them corresponds to in-plane modes (LO, TO, LA, and TA), two quadratic branch (ZA and ZO) corresponds to out-of-plane modes of the structure. Analysis of the phonon spectrum provides a reliable test for the structure stability due to the positive dynamical matrix elements which yields real phonon frequencies. Calculated phonon spectrum of SnC is in good agreement with Ref.\onlinecite{PhysRevB.80.155453}. It is also obvious from the right panel of the Fig. \ref{FIG5} that while the main contribution to the LO, TO and ZO modes comes from the carbon atom, tin atom contributes to acoustic modes, i.e., LA, TA and ZA. Carbon as a light element vibrates with higher frequency in optical branches. The difference between atomic mass for carbon and tin atoms in the unit cell, makes an ideally large energy gap between optical branches. The phonon density of state in the gap is zero. The phonon group velocity which is define as $d\omega/dk$ at long-wavelength limit are 200 m/s, 630 m/s and 1300 m/s for ZA, TA and LA mode, respectively. The phonon velocity is much smaller than graphene and stanene \cite{peng2016low}. 

\begin{figure}[!t]
\includegraphics[width=1.1\linewidth]{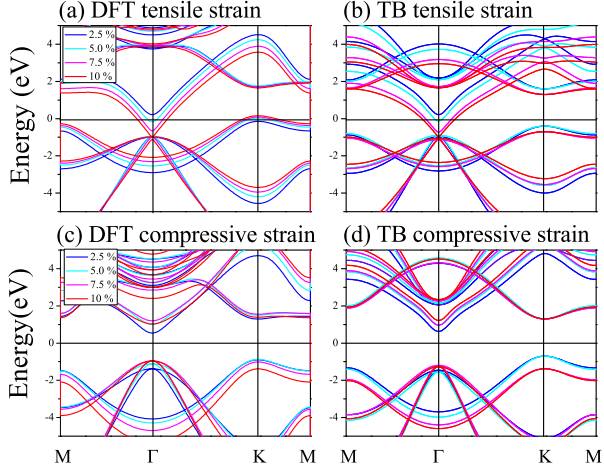}
\caption{The effect of biaxial (a,b) tensile and (c,d) compressive strain on the electronic structure in the DFT-PBE and tight-binding models for SnC.}
\label{STRAIN}
\end{figure}

To modify the electronic band gap we applied a biaxial strain to the hexagonal structure. Since the higher values of strain can induce instability, we apply strain up to the 10$\%$. In the DFT part atomic positions were relaxed in the strained structure. For the tight-binding model the hopping integrals between two atoms are modified according to the generalized form of Harrison’s law \cite{Harrison,TBoykin},

\begin{eqnarray}
t_{i,j,\alpha,\beta}=t_{i,j,\alpha,\beta}^0\times (l/l_0)^\eta, \notag
\end{eqnarray}

where $l_0$ and $l$ are the relaxed and strained atomic bond length. $\eta$ is a parameter which is generally different for each atomic orbital \cite{TBoykin}. Here, for tensile strain and all atomic orbitals $\eta=4$ leads to fair agreement between tight binding and DFT band structure of strained structure. For the compressive strain the $\eta=2.8$ for up to 7\% of strain and $\eta=1.8$ for higher values of strain. To model the indirect to direct band gap transition for compressive strain higher than 7\%, the on site energy of $p_{x,y}$ and $p_z$ atomic orbitals of carbon atoms increases and decreases by amount of 0.5 eV, respectively. The applied tensile or compressive strain changes the atomic distance and consequently the overlap between different atomic orbitals. The net effect of strain is modeled by defining the parameter $\eta$.
Fig. \ref{STRAIN} shows the band structure of strained SnC in the DFT-PBE and tight-binding approximations. Both models predict the variation of band gap with applied strain. The position of VBM and CBM are unchanged by tensile strain \cite{C2JM30915G}. The atomic \textit{s} orbital in the conduction band at the $\Gamma$ point move toward the valence band and fill the energy band gap. The energy band gap at the K point is almost unchanged in the presence of tensile strain. For 10\% of tensile strain the valence and conduction bands cross each other at the $\Gamma$ point. The intrinsic SOC lifts some of degeneracies in band structure and opens energy gap at the $\Gamma$ point for 10\% of tensile strain. The compressive strain slightly modifies the electronic band gap of SnC. For higher values of compressive strain the position of VBM is moved to the $\Gamma$ point and the indirect to direct band gap transition occurs for strain higher than 7.5\%. The proposed tight-binding model correctly predicts the semimetal and the indirect to direct band gap transition for tensile and compressive strain, respectively. 

\begin{figure}[!t]
\includegraphics[width=\linewidth]{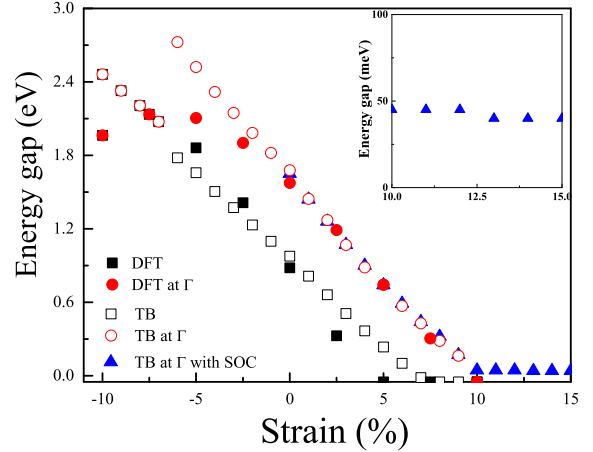}
\caption{The effect of strain on the electronic band gap in the DFT-PBE and tight-binding models.}
\label{GAPFIGURE}
\end{figure}

\begin{figure}[!t]
\includegraphics[width=1.1\linewidth]{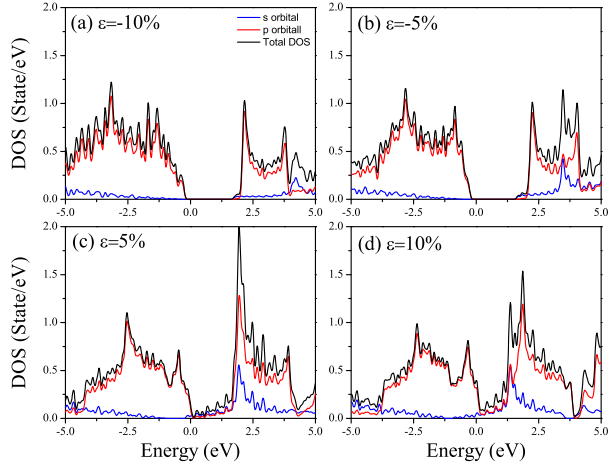}
\caption{The PDOS for a) $\varepsilon=-10$\% b) $\varepsilon=-5$\% c) $\varepsilon=5$\% d) $\varepsilon=10$\%.}
\label{PDOSstrain}
\end{figure}

Fig. \ref{GAPFIGURE} shows the variation of band gap as a function of applied strain in the DFT-PBE and tight-binding approximation. Generally by increasing the lattice constant and atomic distance, the electronic energy band gap decreases and reaches to zero for 7.5\% of tensile strain. The same behavior was reported for 2D nanostructure of BN, AlN and GaN \cite{GapVariation}. The AsN, AsP, SbAs, BiAs \cite{GapVariationdifferent} and single-layer black phosphorus \cite{GapVariationdifferent1} represent different band gap variation with applied compressive and tensile strain. The direct energy gap at the $\Gamma$ point becomes zero for tensile strain higher than 10\%. Although, the SOC has a negligible effect on the electronic structure, it opens approximately $40$ meV energy gap at the $\Gamma$ point for tensile strain higher than 10\%. For higher value of tensile strain, the atomic orbital overlap and the energy gap in the DFT model decreases which cannot be predicted well by the simple tight-binding model. We study the PDOS for strained SnC to understand the atomic orbital contribution on the strained structure. The total PDOS, \textit{s} and \textit{p} decomposition are presented in Fig. \ref{PDOSstrain}. For compressive strain, CBM energy at the $\Gamma$ point increases which opens a gap in band structure with zero PDOS. In the case of tensile strain \textit{s} orbital at the $\Gamma$ and \textit{p} orbital at the K points fill the energy gap and the PDOS is nonzero at the Fermi level. The \textit{s} atomic orbital plays the crucial rule for both compressive and tensile strain.    

\begin{figure}[!t]
\includegraphics[width=1.1\linewidth]{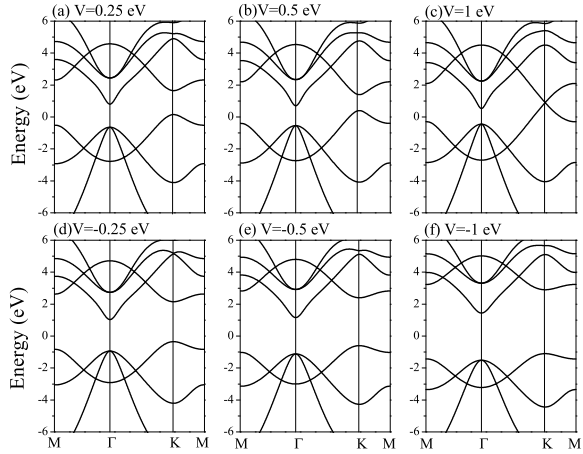}
\caption{The tight-binding electronic band structure of SnC in the presence of (a,b,c) positive and (d,e,f) negative values of staggered potential.}
\label{STP}
\end{figure}

The applied strain modifies the electronic band structure and energy band gap in SnC. Up to here, we consider the free standing SnC. The effect of substrate is modeled with the staggered sub-lattice potential \cite{Sta1,Sta2} which changes the on-site energy of A and B sub-lattices of hexagonal lattice. The staggered potential can be written in the tight-binding model $H^{st}=V\sum_{i,\alpha}\lambda_{i} c_{i,\alpha}^{\dagger }c_{i,\alpha}$, where $V$ is the staggered potential strength and $\lambda_{i}=\pm1$ is different for carbon and tin atoms. The staggered potential may also arises from the perpendicular external electric field. Fig. \ref{STP} shows the band structure of 2D SnC in presence of positive and negative values of staggered potential. The sign and strength of staggered potential is determined by the interaction between SnC and substrate.

For single-layer graphene the staggered potential breaks the space inversion symmetry and opens energy gap in the band structure. In the case of SnC the staggered potential increases or decreases the atomic potential by amount of $V$. Due to difference between on-site energy of carbon and tin atoms the staggered potential may eliminate or increase the on-site energy difference of two atoms. As a result, the electronic gap decreases or increases for positive and negative value of staggered potential. The staggered potential mainly changes the electronic gap at the K point. For $V=1$ eV, the valence and conduction bands cross each other and form a Dirac cone at the K point of the first Brillouin zone. The staggered potential and external strain modify the on-site energies and hopping terms, respectively. The staggered potential changes the sub lattice potential which modifies the gap at the K point. On the other side the external strain modify the hopping between atomic sites which rescale the energy bands in the entire first Brillouin zone and may change the band gap at any k-point.

\begin{figure}[!t]
\includegraphics[width=1.1\linewidth]{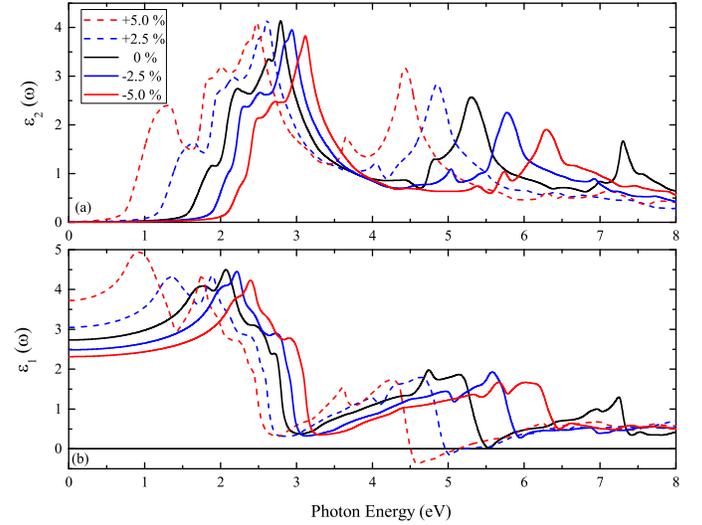}
\caption{The imaginary (a) and real part (b) of dielectric function versus photon energy for the different values of tensile and compressive strains.}
\label{OPTIC}
\end{figure}

To get deeper insight into the electronic structure of considered SnC system, we have also computed behaviour of its dielectric function under the tensile and compressive strain in the low strain regime, i.e., -5\% to 5\%, where the indirect band gap structure of SnC is preserving. Since this function consists of real and imaginary parts with different physical meanings, we consider each part separately and present them in Fig.\ref{OPTIC}. It is well known fact that the imaginary part of the dielectric function ($\varepsilon_{2}$) is an indication of the light absorbing capacity of materials, and it is strongly correlated with the band structure of the material. It can be seen from the Fig.\ref{OPTIC}(a) that onset values (the first step-like increase in the figure) of each curves are compatible with the band gap of the structures in Fig.\ref{STRAIN}. It can be explained by the confinement effect related with bond length shrink or stretching which modifies the electronic structure \cite{PhysRevB.92.165408}. It is also clear from the Fig.\ref{OPTIC}(a) that all the main absorption peaks appear in the visible region (3.26-1.65 eV) of the electromagnetic spectrum. For the tensile strain the peaks are red shifted by amount of $\sim 0.15$ eV from 0 \% to 0.25 \% and 0.25 \% to 0.50 \% which is also reported in some recent works on the optical properties of 2D materials \cite{PhysRevB.87.155304,PhysRevB.92.165408,0022-3727-49-45-455103}. The red shift of the main absorption peaks in the spectrum shows that the band gap is decreased as a consequence of tensile strain. On the other hand, we see that compressive strain yields the blue shift in the spectrum of dielectric function in the SnC monolayer by same amount of shifting, i.e., $\sim 0.15$ eV, related with the increasing of band gap. Moreover the real part of the dielectric function ($\varepsilon_{1}$) is examined in Fig.\ref{OPTIC}(b). The important feature of $\varepsilon_{1}(\omega)$ is the static values in the zero frequency limit, $\varepsilon_{1}(0)$, which shows the dielectric response to the static electric field. While the $\varepsilon_{1} (0)$ takes values 3.72 eV and 3.05 eV (2.31 eV and 2.48 eV) for tensile (compressive) strains 5\% and 2.5\%, respectively, the pristine case has 2.73 eV static dielectric constant. It is also clear that the value of $\varepsilon_{1} (0)$ is inversely proportional with band gap of the strained structures which can be also extracted by the formula, $\varepsilon_{1} (0)=1+(\hbar \omega_{p}/E_{g})$ \cite{PhysRev.128.2093,Guo2013583,dr000000007122}. In other words, static values of  $\varepsilon_{1}$ increase (decrease) with increasing values of tensile (compressive) strain. On the other hand, the negative values of $\varepsilon_{1}$, where the wave decays exponentially in the medium \cite{0034-4885-68-2-R06}, are seen in Fig.\ref{OPTIC}(b). It can be easily claimed that negativity of $\varepsilon_{1}$ is pronounced for only the tensile strain. Furthermore, refractive index of a structure can be calculated in terms of dielectric functions, i.e., $n(\omega)=(\sqrt{\varepsilon_{1}^{2}(\omega)+\varepsilon_{2}^{2}(\omega)}+\varepsilon_{1}(\omega)/2)^{1/2}$. For the pristine SnC, static refractive index, $n(0) $ is found to be 1.65 which is higher than the refractive index of single-layer graphene ($n(0)=1.1$ \cite{rani2014dft}). In the presence of applied strain, the tensile (compressive) strain increases (decreases) the refraction index as $n(0)_{+2.5\%}=1.75$ and $n(0)_{+5.0\%}=1.93$ ($n(0)_{-2.5\%}=1.58$ and $n(0)_{-5.0\%}=1.52$). Eventually, we can easily claim that the optical properties of the SnC system can be tuned by applying tensile and compressive strain for optoelectronic applications which is sensitive to the visible light. Here optical properties are examined in the framework of DFT without local field effects, to observe the many-body effects, i.e., plasmon excitations, excitonic
effects, more sophisticated calculations can be performed by using DFT-beyond models such as HSE, GW, and Bethe-Salpeter equation (BSE) \cite{WCMS:WCMS1252}.

\section{Conclusion}
In summary, we investigate that the electronic structure and optical properties of strained SnC using the DFT and tight-binding models. The appropriate tight-binding parameters for relaxed and strained SnC are presented. The positive phonon spectrum confirms the stability of 2D structure and the electronic PDOS determines the contribution of different atomic orbitals in the electronic properties of SnC. We conclude that the phonon frequency gap is related to the different atomic mass of carbon and tin atoms in the unit-cell. We also analysis the Mulliken charge population which shows the covalent nature of carbon and tin bond in SnC. Our results show the importance of atomic \textit{s} orbital in relaxed and strained SnC. Moreover, for SnC the applied tensile and compressive strain may yield semiconductor to semimetal and the indirect to direct band gap transition, respectively. We also analyze the effect of substrate on the electronic structure of SnC modeled by tight-binding approach within the staggered potential term. Finally, we study the dielectric function of SnC in the presence of strain and show that SnC is a promising material for the potential optoelectronic applications by strain engineering.

\section{Acknowledgments}
The authors thank A. N. Rudenko for fruitful discussions.

\section*{References}
\bibliography{referans}
\bibliographystyle{apsrev4-1}

\end{document}